\documentclass[journal,twoside,web]{ieeecolor}
\usepackage{tmi}
\usepackage{cite}
\usepackage{amsmath,amssymb,amsfonts}
\usepackage{algorithmic}
\usepackage{graphicx}
\usepackage{textcomp}
\usepackage{url}
\usepackage{tabularx}
\usepackage{multicol}
\usepackage{multirow}
\usepackage[labelsep=period]{caption}
\usepackage{booktabs}
\usepackage{url}
\usepackage{capt-of}
\usepackage{adjustbox}
\usepackage{dblfloatfix}
\usepackage{subfigure}

\def\BibTeX{{\rm B\kern-.05em{\sc i\kern-.025em b}\kern-.08em
    T\kern-.1667em\lower.7ex\hbox{E}\kern-.125emX}}
\begin{document}
\title{Multi-scale Feature Enhancement in Multi-task Learning for Medical Image Analysis}
\author{Phuoc-Nguyen Bui, Duc-Tai Le, Junghyun Bum, Hyunseung Choo \IEEEmembership{Member, IEEE}
}

\maketitle

\begin{abstract}
Traditional deep learning methods in medical imaging often focus solely on segmentation or classification, limiting their ability to leverage shared information. Multi-task learning (MTL) addresses this by combining both tasks through shared representations but often struggles to balance local spatial features for segmentation and global semantic features for classification, leading to suboptimal performance. In this paper, we propose a simple yet effective UNet-based MTL model, where features extracted by the encoder are used to predict classification labels, while the decoder produces the segmentation mask. The model introduces an advanced encoder incorporating a novel ResFormer block that integrates local context from convolutional feature extraction with long-range dependencies modeled by the Transformer. This design captures broader contextual relationships and fine-grained details, improving classification and segmentation accuracy. To enhance classification performance, multi-scale features from different encoder levels are combined to leverage the hierarchical representation of the input image. For segmentation, the features passed to the decoder via skip connections are refined using a novel dilated feature enhancement (DFE) module, which captures information at different scales through three parallel convolution branches with varying dilation rates. This allows the decoder to detect lesions of varying sizes with greater accuracy. Experimental results across multiple medical datasets confirm the superior performance of our model in both segmentation and classification tasks, compared to state-of-the-art single-task and multi-task learning methods. The code will be available at \url{https://github.com/nguyenpbui/ResFormer}.
\end{abstract}

\section{Introduction}
\label{sec:introduction}
Medical image classification and segmentation are two fundamental tasks in the field of medical imaging, both essential for accurate diagnosis and effective treatment planning. Classification assigns diagnostic labels based on the overall pathology in an image, while segmentation involves identifying and delineating specific regions of interest, such as tumors or lesions. Traditionally, deep learning models have been designed to tackle these tasks separately through single-task learning (STL) approaches. While effective in specific use cases, these methods suffer from a critical limitation: they fail to leverage the inherent connection between classification and segmentation. Because identifying abnormal regions directly affects the overall diagnostic outcome, treating these tasks independently overlooks valuable shared information. This task isolation limits the model's ability to fully exploit the rich feature representations in medical images, resulting in suboptimal performance for both segmentation and classification.

\begin{figure*}[!ht]
    \centering
    \includegraphics[width=0.85\textwidth]{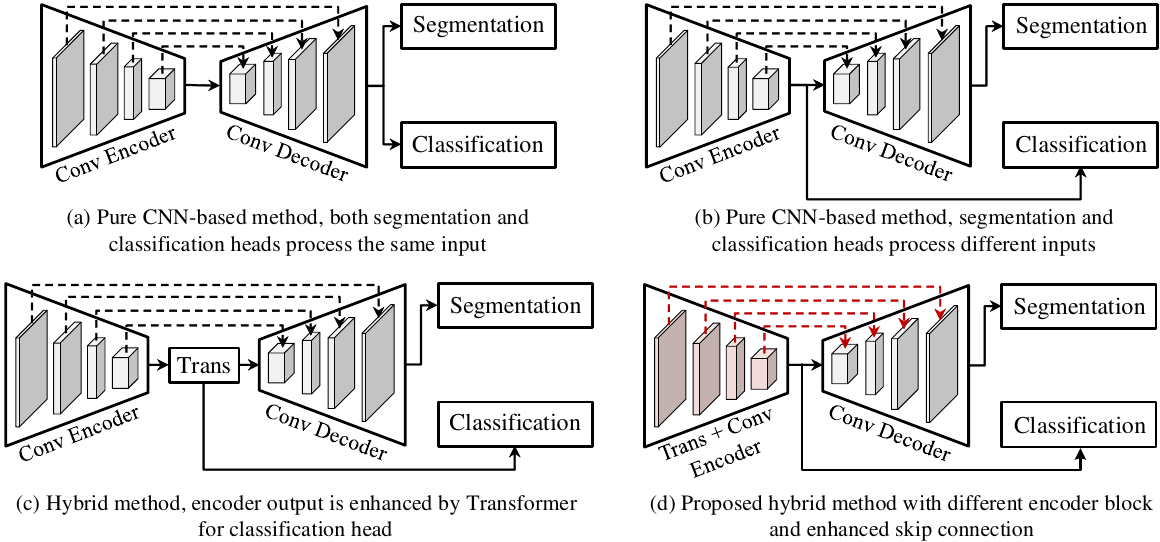}
    \caption{Evolution of multi-task learning methods in medical image classification and segmentation. Conv: Convolutional, Trans: Transformer.}
    \label{fig: introduction}
\end{figure*}

Multi-task learning (MTL) has emerged as a promising solution to address these challenges by harnessing shared information across interrelated tasks within a single model, thereby improving the performance of both classification and segmentation. Early studies on MTL in medical image processing \cite{he2020multi, gende2022end, hervella2022end} primarily employed convolutional neural networks (CNNs) within encoder-decoder architectures, with the U-Net \cite{ronneberger2015u} being a widely adopted model. In these frameworks, the classification and segmentation heads either process the same features from the encoder or use different inputs, as shown in Fig. \ref{fig: introduction} (a-b), respectively. Recently, hybrid methods that combine CNNs with Transformers have revolutionized MTL in medical image analysis. By leveraging the self-attention mechanism of Transformers \cite{vaswani2017attention}, these models can effectively capture long-range dependencies and contextual relationships \cite{cheng2022fully, tang2023transformer}, as depicted in Fig. \ref{fig: introduction} (c).

Despite these advancements, both CNN-based and hybrid methods still encounter certain limitations. While CNN-based approaches are effective at capturing local features, they often struggle to adequately capture long-range dependencies in medical images owing to the significant variability in lesion shapes and sizes \cite{lin2017feature}. This issue stems from the inherent bias of small convolutional kernels toward local feature extraction, which limits the network's ability to effectively encode global context \cite{yu2015multi} effectively. On the other hand, hybrid approaches that combine CNNs and Transformers aim to harness the strengths of both local and global feature extraction, showing promise in capturing global relationships through the attention mechanism, thereby enhancing overall performance. However, existing methods may be prone to multi-scale information loss during the fusion of CNN and Transformer layers since they typically apply the Transformer only after the last stage of the encoder branch, as shown in Fig. \ref{fig: introduction}(c). This potential loss of information impedes the model's overall effectiveness, highlighting the need for further refinement and innovation in integrating these two architectures.

This paper introduces a hybrid MTL method for lesion segmentation and classification in medical images designed to overcome these drawbacks. In particular, we propose a novel encoder building block, called ResFormer, which integrates the strengths of convolutional neural networks and the Transformer in either sequential or parallel manners, facilitating better feature extraction for both tasks. For classification, we leverage features from multiple encoder levels to predict image-level labels, enabling the model to capture a comprehensive and hierarchical representation of the input data. To further improve segmentation performance, we enhance the encoder's features before feeding them into the segmentation decoder using a novel dilated feature enhancement (DFE) module. The DFE module comprises three parallel convolution branches with different dilation rates, effectively capturing the multi-scale contextual information, and thus facilitating the detection of lesions from varying sizes. Extensive experiments on the RETOUCH and ISIC 2017 datasets demonstrate the superior performance of our proposed method compared to existing STL and MTL approaches, highlighting its potential to advance medical image segmentation and classification. The key contributions of this paper are summarized as follows:

$\bullet$ We propose a simple yet effective UNet-based MTL method for lesion segmentation and classification, combining the local feature extraction of CNNs with the long-range dependency capture of Transformers in a unified block, called ResFormer, enhancing feature representation and performance.

$\bullet$ We introduce a dilated feature enhancement module specifically designed for the segmentation decoder. This module employs dilated convolutional layers with various dilation rates to capture information at different scales, improving lesion segmentation in medical images under various scenarios of medical images.

$\bullet$ Extensive experiments on the eyes and skin datasets confirm that our method significantly outperforms current state-of-the-art approaches in both segmentation and classification tasks. The consistent performance across these diverse datasets underscores the robustness and generalizability of our approach in addressing various medical imaging challenges.

The remainder of this article is organized as follows: Section II provides a comprehensive review of related work. Section III delves into the details of the proposed method. Section IV outlines the datasets used, implementation specifics, and evaluation metrics. Section V presents a thorough performance evaluation. Finally, Section VI concludes the article with a summary of findings and suggestions for future research.

\section{Related work}
\subsection{Multi-task learning in medical imaging}
Multi-task learning has been employed in medical imaging to enhance performance across various tasks by leveraging shared information and representations. MTL approaches can be classified into two categories: CNN-based \cite{he2020multi, gende2022end, hervella2022end, wang2023efficient, he2023joint, diao2023classification, ling2023mtanet} and hybrid \cite{tang2023transformer, marcus2023concurrent, fan2023joint} methods. We elaborate on these categories below.

\textbf{CNN-based methods}: Researchers have extended the conventional U-Net \cite{ronneberger2015u} to address both lesion segmentation and disease classification tasks simultaneously. For example, He et al. \cite{he2020multi} developed an encoder-decoder network for organ segmentation and multi-label classification in CT images.  In their approach, the primary task focused on pixel-wise organ segmentation, while the auxiliary task involved image-level classification of multiple organs in the same CT images. Similarly, Gende et al. \cite{gende2022end} proposed an end-to-end, fully automated method for concurrently segmenting relevant structures and detecting disease symptoms in OCT images. More recently, Wang et al. \cite{wang2023efficient} developed a two-stage framework that leverages class activation maps from a polyp classification model to enhance segmentation performance. Meanwhile, Ling et al. \cite{ling2023mtanet} introduced the multi-task attention network (MTANet), a unified framework designed to efficiently perform object classification while generating high-quality segmentation masks for medical objects. This model incorporates a novel reverse addition attention module to enhance feature extraction and improve segmentation precision. As aforementioned, these methods struggle to capture long-range dependencies in medical images, which are crucial for precise segmentation.

\textbf{Hybrid methods}: Given the promising performance of Transformers, hybrid approaches aim to leverage the strengths of Transformers to achieve superior performance. For example, Cheng et al. \cite{cheng2022fully} proposed a transformer-based multi-task learning framework for simultaneous glioma segmentation and isocitrate dehydrogenase (IDH) classification, incorporating a transformer block between the encoder and decoder to extract global information. Similarly, Tang et al. \cite{tang2023transformer} introduced a multi-task network (TransMT-Net) to classify and segment gastrointestinal tract endoscopic images. Marcus et al. \cite{marcus2023concurrent} recently proposed a transformer-based network for simultaneous segmentation and age estimation of cerebral ischemic lesions. The model incorporates gated positional self-attention and CT-specific data augmentation to effectively capture long-range dependencies, particularly in scenarios with limited training data. However, these methods only employ transformers at the end of the encoder branch, which compromises the retention of multi-scale features, resulting in suboptimal performance owing to the loss of critical spatial information.

\subsection{Single-task learning in medical imaging}
Single-task learning (STL) methods have been crucial in medical image analysis, particularly for 2D imaging tasks such as classification and segmentation. These methods optimize models for specific tasks, leading to highly specialized and effective solutions.

\textbf{Image classification}: Convolutional neural networks (CNNs) have historically demonstrated exceptional performance, with models such as ResNet \cite{he2016deep} and InceptionNet \cite{szegedy2016rethinking} being notable examples recognized for their ability to extract hierarchical features and improve classification accuracy. These models have been effectively applied to classify various diseases in 2D medical images including skin lesions in dermoscopy images and abnormalities in OCT images \cite{kermany2018identifying, sunija2021octnet, thomas2021novel}. Recently, transformers have recently been adapted for medical image classification, offering several advantages over traditional CNNs such as the ability to capture long-range dependencies and contextual information. The vision transformer (ViT) \cite{dosovitskiy2010image} is a pioneering model in this field, treating images as sequences of patches and processing them with transformer encoders. ViT has shown competitive performance in medical image classification tasks by effectively capturing global context. Meanwhile, the swin transformer (SwinT) \cite{liu2021swin} uses a hierarchical structure and shifted windows to compute self-attention, which enables it to capture both local and global features and thereby improving classification accuracy.

\textbf{Image segmentation}: The U-Net architecture \cite{ronneberger2015u} has been a cornerstone in this field, utilizing an encoder-decoder structure with skip connections to capture both local and global features effectively. Owing to its robustness and accuracy, U-Net has been widely adopted and extended acroess various medical imaging applications \cite{rasti2022retifluidnet, xing2022multi, he2022intra, he2022structure, yu2023loss, hassan2021joint, diao2023classification}. For example, Xing et al. \cite{xing2022multi} improved the U-Net architecture by integrating spatial pyramid pooling modules to extract multi-scale objects for fluid segmentation in OCT images. Another considerable advancement is the DeepLab series \cite{chen2018encoder}, which incorporates atrous (dilated) convolutions to capture multi-scale contextual information, enhancing segmentation accuracy in complex 2D medical images such as retinal scans and histopathology slides. More recently, transformer-based models such as TransUNet \cite{chen2021transunet}, Swin-Unet \cite{cao2022swin}, HiFormer \cite{heidari2023hiformer} and MsTGANet \cite{wang2021mstganet} have been introduced, combining CNNs and transformers to leverage both local and global dependencies for improved segmentation performance.

\section{Methodology}

\begin{figure*}[!t]
    \centering
    \includegraphics[width=0.85\textwidth]{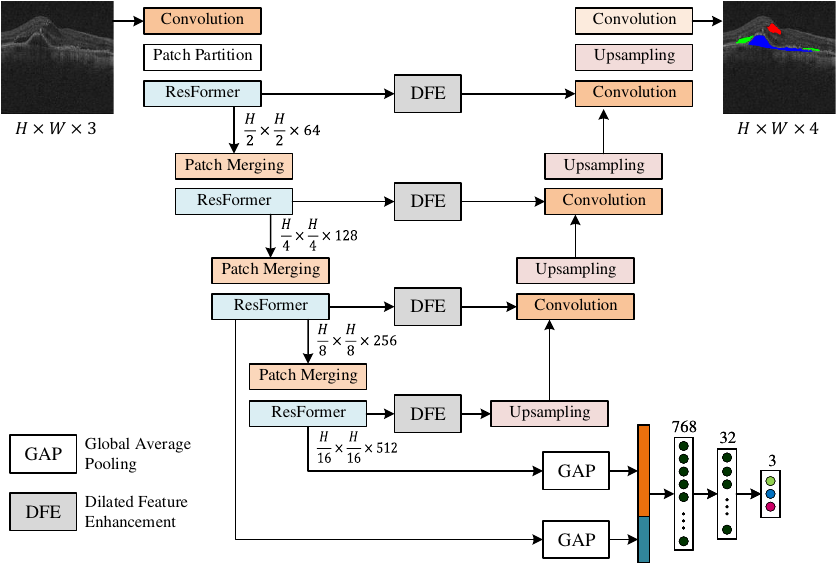}
    \caption{The overview of multi-task learning method for medical image classification and segmentation. The proposed method utilizes the U-Net architecture with a shared encoder and two dedicated decoders for the classification and segmentation tasks, respectively.}
    \label{fig: overall_architecture}
\end{figure*}

The proposed method utilizes a U-Net architecture \cite{ronneberger2015u}, as illustrated in Figure \ref{fig: overall_architecture}. Given an input image in the size of $H \times W \times 3$, the encoder extracts hierarchical semantic information using a novel ResFormer block as the network goes from levels 1 to 4. In the classification head, we integrate the hierarchical features from multiple encoder levels to predict image-level labels. For segmentation, we incorporate a dilated feature enhancement (DFE) module at each encoder level to enhance the corresponding features before passing them to the decoder block. The decoder's output is then used to produce pixel-level labels. We provide detailed explanations of each component in the following sections. 

\subsection{Encoder}
\begin{figure}[!t]
    \centering
    \includegraphics[width=\columnwidth]{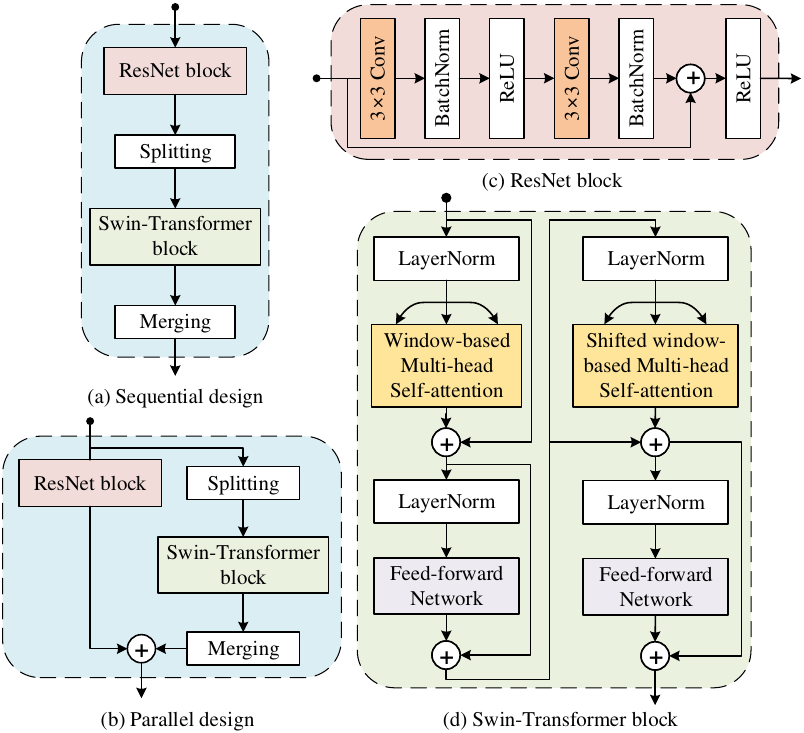}
    \caption{Two designs of the proposed ResFormer block which combines ResNet \cite{he2016deep} and Swin-Transformer \cite{liu2021swin} blocks.}
    \label{fig: resformer-resnet-swint}
\end{figure}

Given an input image $I$ with dimensions $H \times W \times 3$, where $H$ and $W$ represent the spatial height and width, respectively, and $3$ denotes the number of color channels. The proposed encoder begins by employing a convolutional followed by a Max-Pooling layer to establish the initial feature map, $F_0$ at the size of ${\frac{H}{2} \times \frac{W}{2} \times 64}$, which will be used as the input for the ResFormer block. The proposed encoder block leverages the strengths of both CNN and Transformer by connecting a convolutional block from the ResNet34 architecture \cite{he2016deep} with a Swin-Transformer block \cite{liu2021swin} in either a sequential or a parallel manner, as respectively illustrated in Fig. \ref{fig: resformer-resnet-swint} (a) and (b). These designs allow the ResFormer block to effectively capture the local and global features for improved representation learning.

\textbf{ResNet34 block:} It is well-known for employing residual connections to mitigate the vanishing gradient problem, thereby enabling efficient neural network training, particularly in very deep architectures. As illustrated in Fig. \ref{fig: resformer-resnet-swint} (c), the ResNet block comprises a series of 3 $\times$ 3 convolutional ($\text{Conv}_{3\times3}$) layers, each followed by batch normalization (BN) and ReLU activation. These residual connections allow the input to bypass certain layers and be directly added to the output, thereby preserving essential information during training and improving convergence. Let $F_i$ denote the input feature maps of the encoder level $i$ and $F^{\text{R}}_i$ be the output feature maps of the ResNet block at the level. The detailed operation can be formulated as follows:
\[
F^{\text{R}}_i = \text{ReLU}(F_i+F'_i)
\]
where $F'_i = \text{BN(Conv}_{3\times3}\text{(ReLU(BN(Conv}_{3\times3}(F_i)))))$.

\textbf{Swin-Transformer block:} It is upgraded from the standard transformer block \cite{vaswani2017attention} by replacing the conventional multi-head self-attention (MSA) module with a window-based (W-MSA) and a shifted window-based (SW-MSA) mechanisms while maintaining the other layers unchanged. In each Swin Transformer block, W-MSA and SW-MSA modules are followed by feed-forward networks (FFN). Each MSA and FFN module is preceded by a LayerNorm (LN) layer, and residual connections are applied after each of the modules, as shown in Fig. \ref{fig: resformer-resnet-swint} (d). Let the input feature map of the $i^{th}$ Swin-Transformer block be denoted as $\tilde{F_i}$. This input is first normalized and then processed by the W-MSA module to produce the feature map $F^\text{W}_i$, formulated as follows:
\[
F^\text{W}_i = \text{W-MSA}(\text{LN}(\tilde{F_i})) + \tilde{F_i}
\]

\begin{figure*}[!hb]
    \centering
    \includegraphics[width=0.9\textwidth]{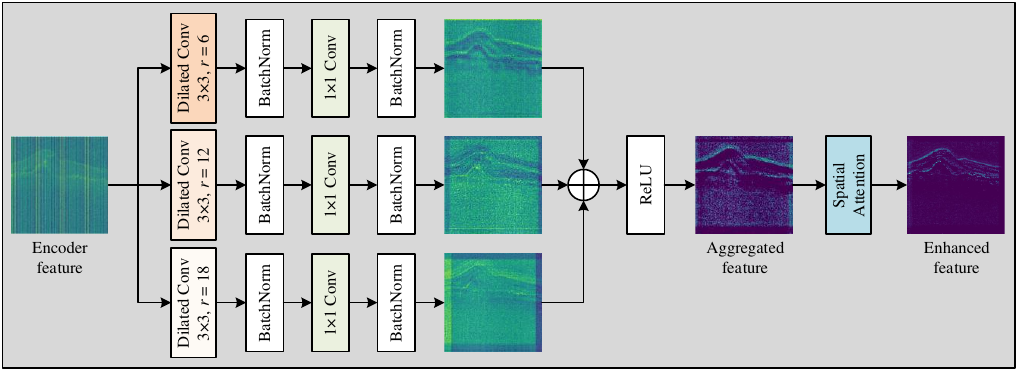}
    \caption{The proposed dilated feature enhancement (DFE) module.}
    \label{fig: dfe}
\end{figure*}

Subsequently, the feature map $F^\text{W}_i$ is processed by the SW-MSA module after a pair of LN and FFN. The output $F^\text{SW}_i$ of the SW-MSA is generated as follows:
\[
F^{\text{SW}}_i = \text{SW-MSA}(\text{LN}(\tilde{F'_i})) + \tilde{F'_i}
\]
where  $\tilde{F'_i}= \text{FFN}(\text{LN}(F^\text{W}_i)) + F^\text{W}_i$ is the intermediate feature map. The output $F^{\text{T}}_i$ of the $i^{th}$ Swin-Transformer block is computed by passing the $F^{\text{SW}}_i$ through another pair of LN and FFN, which is described as follows:
\[
F^{\text{T}}_i = \text{FFN}(\text{LN}(F^{\text{SW}}_i) + F^{\text{SW}}_i
\]

The Swin-Transformer block is configured with 2, 4, 8, and 16 attention heads across four levels from 1 to 4, respectively, enabling it to capture complex feature representations across multiple scales. Meanwhile, the FFN consists of two linear projection (Linear) layers, and a GELU activation function is applied between them. Consistent with \cite{liu2021swin}, the expansion ratio of the Linear layers is set to 4 in our experiments.

\textbf{Sequential ResFormer}: In this design, the input feature map at each encoder level is first fed to the ResNet34 block to capture the local contextual information and inject a strong inductive bias. We then split the ResNet34 block's output into patches via the Splitting operation before feeding it to the Swin-Transformer block, i.e. $\tilde{F_i}= F^{\text{R}}_i$, to learn the long-range dependencies and global information. Finally, the output of the ResFormer block is reconstructed into 2D feature maps using a Merging operation on the Swin-Transformer block's output, effectively combining both local and global features for enhanced representation learning. The output feature map at the encoder level $i$ can be formulated as follows:
\[
\hat{F_i} = F^{\text{T}}_i
\]
\textbf{Parallel ResFormer}: In this design, the ResNet34 and Swin-Transformer blocks operate in parallel. Unlike the sequential design, where the Swin-Transformer processes the output of the ResNet34 block, both blocks receive the same input feature map directly. This parallel structure allows each block to independently extract complementary features, i.e. local spatial details from the ResNet34 block and global contextual information from the Swin-Transformer. The final output of the ResFormer block is then generated by fusing the feature maps from both branches through element-wise summation, resulting in a richer and more comprehensive feature representation. Formally, the output feature map at level $i$ is:
\[
\hat{F_i} = F^{\text{R}}_i + F^{\text{T}}_i
\]
The encoder generates hierarchical feature maps at four levels, with each level reducing the spatial resolution to 1/2, 1/4, 1/8, and 1/16 of the original resolution, respectively. Following the approach in \cite{liu2021swin}, we incorporate a PatchMerging operation between two adjacent encoder levels for efficient token reduction and dimensional expansion. This is accomplished through a 3$\times$3 convolutional layer with a stride of 2.

\subsection{Decoder}
In our method, the decoder refines and enhances the features extracted by the encoder, integrating multi-scale information to produce precise segmentation maps. Unlike the conventional U-Net which utilizes skip connections, we introduce a dilated feature enhancement (DFE) module to capture multi-scale feature representation at each encoder level, incorporating both fine- and coarse-grained details. 

\textbf{DFE module:} As illustrated in Figure \ref{fig: dfe}, the DFE module refines the encoder feature map, $\hat{F_i}$, through three parallel dilated $3\times 3$ convolutional ($\text{Conv}^d_{3\times3}$) branches, each with different dilation rates $d=\{6, 12, 18\}$. In each branch, the dilated convolution is followed by a BN layer, a 1$\times$1 convolutional ($\text{Conv}_{1\times1}$) layer, and another BN layer, ensuring efficient feature extraction and normalization. The outputs from these branches, i.e. $\hat{F}^6_i$, $\hat{F}^{12}_i$, and $\hat{F}^{18}_i$, are then combined through summation, activated by a ReLU function, and processed by a spatial attention (SA) module \cite{woo2018cbam} to ignore the irrelevant areas. This process results in an enhanced feature, denoted as $\hat{F^{\text{e}}_i}$, for the decoder. The operations of the DFE module are expressed as follows:
\begin{equation*}
\begin{split}
\hat{F}^6_i & = \text{BN(Conv}_{1\times1}\text{(BN(DConv}_{3\times3}^6(\hat{F_i}))))) \\
\hat{F}^{12}_i & = \text{BN(Conv}_{1\times1}\text{(BN(DConv}_{3\times3}^{12}(\hat{F_i}))))) \\
\hat{F}^{18}_i & = \text{BN(Conv}_{1\times1}\text{(BN(DConv}_{3\times3}^{18}(\hat{F_i}))))) \\
\hat{F}^{\text{e}}_i & = \text{SA(ReLU(}\hat{F}^{6}_i + \hat{F}^{12}_i + \hat{F}^{18}_i))
\end{split}
\end{equation*}

At each level in the decoder except the first one, the output of the DFE module, $\hat{F^{\text{e}}_i}$, is concatenated (Concat) channel-wise with the upsampled feature from the preceding decoder, $F^{\text{d}}_{i}$. This concatenated feature is then processed by a 3$\times$3 convolution, followed by a batch normalization layer and a ReLU activation. Finally, we upsample (Upsample) the feature to produce the decoder output at level $i$, $\hat{F}^{\text{d}}_i$, $i=\{2, 3, 4\}$. Given the $\hat{F}^{\text{d}}_1 = \text{Upsample}(\hat{F}^{\text{e}}_4)$. The operations are detailed as follows:
\[
\hat{F}^{\text{d}}_i = \text{Upsample(ReLU(BN(Conv}_{3\times3}(\text{Concat}(\hat{F}^{\text{e}}_{5-i}, F^{\text{d}}_{i})
\]

\subsection{\textit{Output heads and loss functions}}

\textbf{Classification head}: To leverage the advantages of multi-scale learning, we apply global average pooling (GAP) on the features map of the last two encoder levels, resulting in 256- and 512-dimensional vectors, respectively. These vectors are then channel-wise concatenated (Concat) and fed through a simple network consisting of two fully connected (FC) layers. This network yields the image-level output, $\hat{y}_{\text{cls}}$, an $n$-dimensional vector representing the $n$ types of the lesion, as follows:
\[
\hat{y}_{\text{cls}} = \text{FC(FC(Concat(GAP(}\hat{F}_3), \text{GAP(}\hat{F}_4))))
\]

\textbf{Segmentation head:} In the final decoder level, a 1$\times$1 convolution followed by a softmax activation function (Softmax) is applied on the decode features to produce a $n+1$ output, corresponding to $n$ types of lesion and the background label. Let the output of the segmentation head be denoted as $\hat{y}_{\text{seg}}$. The detailed process is described as follows:
\[
\hat{y}_{\text{seg}} = \text{Softmax(Conv}_{1\times1}(\hat{F}^{\text{d}}_4)) \\
\]

\textbf{Loss function:} During training, the supervised loss $\mathcal{L}$ is defined as a weighted combination of two components: the weighted multi-label loss $\mathcal{L_{\text{wce}}}$ for classification task and the Dice loss $\mathcal{L_{\text{dice}}}$ for segmentation task. The combined loss function is formulated as follows:
\[
\mathcal{L} = \lambda_1 * \mathcal{L_{\text{wce}}}(\hat{y}_{\text{cls}}, y_{\text{cls}}) + \lambda_2 * \mathcal{L_{\text{dice}}} (\hat{y}_{\text{seg}}, y_{\text{seg}})
\]
where $y_{\text{cls}}$ and $y_{\text{seg}}$ are the ground truth of classification and segmentation tasks, respectively. The weight parameters $\lambda_1$ and $\lambda_2$ are used to balance the contribution of each loss function. In our experiments, these parameters are set to 0.25 and 1, respectively.

\section{Experiments}
\textbf{Datasets:} The RETOUCH dataset comprises 112 macula-centered OCT volumes, with 70 volumes allocated for training and 42 for testing. These OCT volumes were captured using spectral-domain SD-OCT systems from three prominent vendors: Cirrus HD-OCT (Zeiss Meditec), Spectralis (Heidelberg Engineering), and T-1000/T2000 (Topcon). Each Cirrus and Topcon OCT volume includes 128 B-scans, while those acquired using the Spectralis device consist of 49 B-scans, offering a varied range of imaging resolutions and perspectives across different devices. The labels for the test set are unavailable so we split all volumes into images and divide the original training set into a training set and a test set for the experiments as summarized in Table \ref{dataset: OCT}. 

\begin{table}[!ht]\small
\caption{RETOUCH dataset summary.}
\setlength{\tabcolsep}{3pt}
\centering
\begin{tabular}{|c|c|c|c|c|}
\hline
Machine & \# of volumes & Image size & Training & Testing\\
\hline
Cirrus & 24 & 512 $\times$ 1024 & 2560 & 512 \\
\hline
Spectralis & 24 & 512 $\times$ 496 & 980 & 198 \\
\hline
Topcon & 22 & 512 $\times$ 885 & 2176 & 512 \\
\hline
\end{tabular} 
\label{dataset: OCT}
\end{table}

The second dataset is from the International Skin Imaging Collaboration (ISIC) 2017 challenge. It includes high-resolution images, typically around 600 $\times$ 450 pixels, annotated with segmentation masks and diagnostic labels for various lesion types, including melanoma, nevus, and seborrheic keratosis. The dataset is partitioned into a training set of 2,000 images, a validation set of 150 images, and a testing set of 600 images, each annotated with segmentation masks and diagnostic labels for three skin lesions. This dataset is widely used to advance the accuracy and effectiveness of automated skin lesion analysis techniques.

\textbf{Implementation details:} To mitigate the overfitting and improve the model's generalization capability, we first resize the images and then apply data augmentation techniques such as random rotation, and horizontal/vertical flip. For the OCT dataset, the input size is set to 512 $\times$ 512 to preserve the details. For the ISIC 2017 dataset, all images are resized to 224 $\times$ 224, following settings in other methods. The proposed method is implemented using the PyTorch framework with randomly initialized weights. The model is trained with a batch size of 8, a learning rate of 0.003, and 100 epochs. The AdamW optimizer is employed, with momentum coefficients set to (0.9, 0.999) and a weight decay of 0.0001. The checkpoint with lowest validation loss is used for testing. To ensure robust and reproducible results, all experiments are conducted using five different random seeds, and the mean values are computed for evaluation.

\textbf{Classification evaluation metrics}: To obtain a more accurate and balanced assessment of the model's overall performance, we calculate the micro-average ($\mu$-average) for each evaluation metric, including accuracy, sensitivity, and specificity, as follows:
\[ \mu \text{Acc} = \frac{\sum_{i=1}^C(TP_i + TN_i)}{\sum_{i=1}^C(TP_i + FP_i + TN_i + FN_i)} \]
\[ \mu \text{Sen} = \frac{\sum_{i=1}^C TP_i}{\sum_{i=1}^C(TP_i + FN_i)} \]
\[ \mu \text{Spe} = \frac{\sum_{i=1}^C TN_i}{\sum_{i=1}^C(TN_i + FP_i)} \] 

 where C denotes the number of classes while TP, TN, FP, and FN represent the counts of true positives, true negatives, false positives, and false negatives, respectively. 

\textbf{Segmentation evaluation metrics}: 
Dice similarity coefficient (DSC): Twice the number of pixels in the overlap area divided by the total pixels in the PR mask and the GT. 
\[ DSC = \frac{2\big|GT \cap PR\big|}{\big|GT\big| + \big|PR\big|} =\frac{2TP}{2TP + FP + FN} \]
    
Jaccard Index (JI): The number of pixels in the overlap area over those in the union area between the PR mask and the GT.
 \[ JI = \frac{\big|GT \cap PR\big|}{\big|GT \cup PR\big|} =\frac{TP}{TP + FP + FN} \]

\section{Performance Evaluation}

\subsection{Comparison with existing MTL methods}
To demonstrate the effectiveness of the proposed method, we compare it with both CNN-based methods, i.e. He et al. \cite{he2020multi}, Hervella et al. \cite{hervella2022end}, and Gende et al. \cite{gende2022end}, and hybrid method Tang et al. \cite{tang2023transformer}. Because these methods were originally developed for different datasets, we adhere strictly to their papers or use open-source codes to reproduce the results, ensuring a fair comparison.

\begin{figure}[!h]
    \centering
    \includegraphics[width=\linewidth]{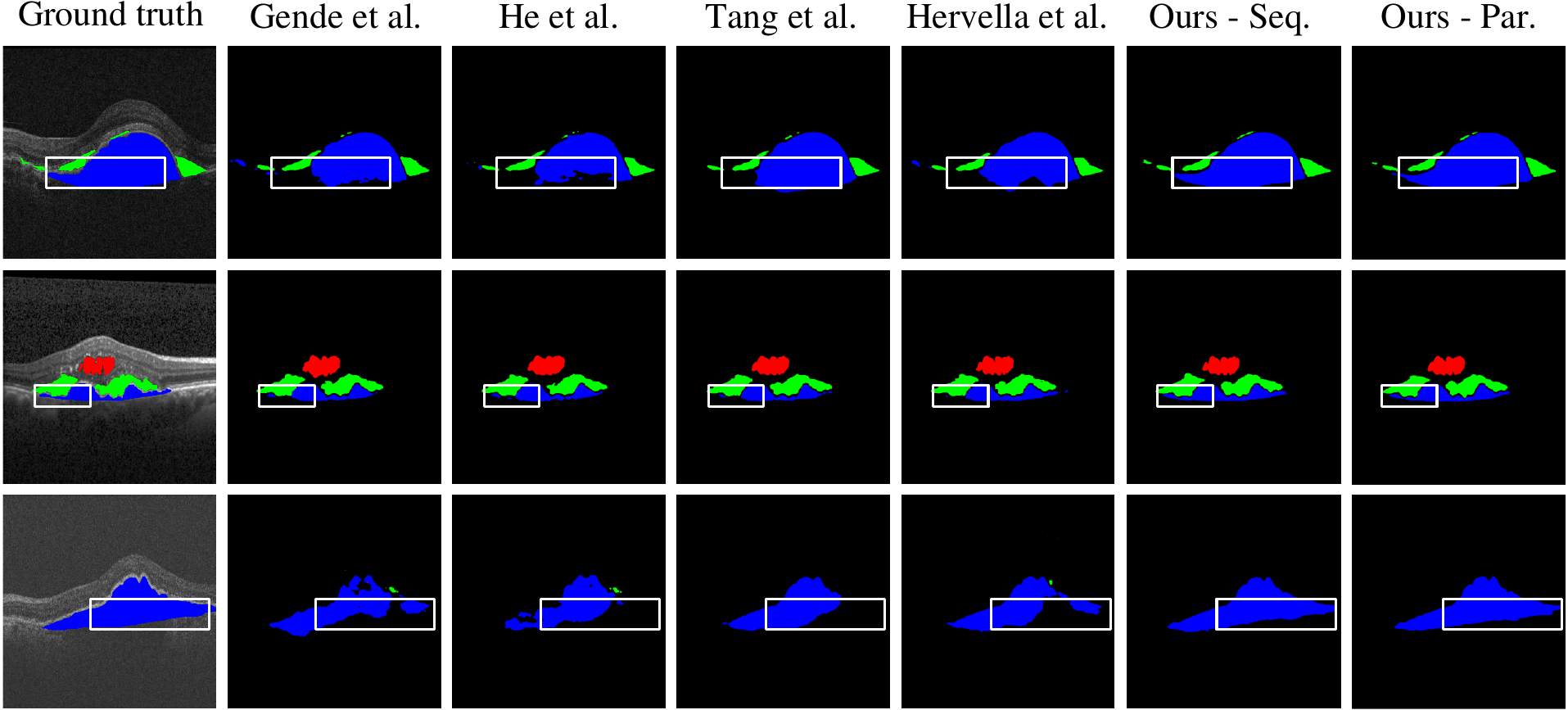}
    \caption{Segmentation visualization of MTL methods on the RETOUCH dataset. Red, green, and blue indicate intra-retinal fluid, sub-retinal fluid, and pigment epithelial detachment, respectively.}
    \label{Fig: Results_MTL_RETOUCH}
\end{figure}

\textbf{Results on the RETOUCH dataset:} In the classification task, our method, including both sequential and parallel variants, exhibits superior performance compared to other MTL approaches, as evidenced by the results in Table \ref{tab: MTL-Cls-RETOUCH}. Specifically, the parallel variant of our method achieves the highest micro-average accuracy ($\mu$Acc) of 99.0\% on the Cirrus machine, while the sequential variant achieves the best $\mu$Acc of 97.8\% and 98.0\% on Spectralis and Topcon machines, respectively. The quantitative results for retinal fluid segmentation are presented in Table \ref{tab: MTL-Seg-RETOUCH}. Our parallel variant achieves remarkable JI and DSC scores, significantly outperforming the previous state-of-the-art method. Specifically, we observe improvements of 2.8\%, 3.3\%, and 3.6\% in DSC for the Cirrus, Spectralis, and Topcon machines, respectively. The superior performance of our method can be attributed to the novel ResFormer block, which efficiently integrates both spatial local information and long-range dependencies within a unified structure. Additionally, the qualitative results of our method, compared to others, are illustrated in Figure \ref{Fig: Results_MTL_RETOUCH}. The areas highlighted by white boxes show our method's consistent superiority across large and small fluid scenarios.

\begin{table}[!h]\small
\caption{Classification performance of MTL methods on the RETOUCH dataset. The best and second-best results are respectively highlighted in \textcolor{red}{\textbf{red}}, \textcolor{blue}{\textbf{blue}}.}
\setlength{\tabcolsep}{2.5pt}
\centering
\begin{tabular}{l|ccc|ccc|ccc}
\toprule
\multirow{2}{*}{Method} & \multicolumn{3}{c|}{Cirrus} & \multicolumn{3}{c|}{Spectralis} & \multicolumn{3}{c}{Topcon} \\
 & Acc & Sen & Spe & Acc & Sen & Spe & Acc & Sen & Spe \\
\midrule
He et al. & 98.4 & 97.3 & 98.9 & 97.0 & 95.4 & 97.8 & 97.8 & 96.5 & \textcolor{blue}{\textbf{98.3}} \\
Gende et al. & 98.7 & 98.0 & 99.0 & \textcolor{blue}{\textbf{97.7}} & \textcolor{blue}{\textbf{96.1}} & \textcolor{blue}{\textbf{98.5}} & 97.4 & 95.8 & 98.1 \\
Hervella et al.  & 96.9 & 95.2 & 97.6 & 94.3 & 91.0 & 96.0 & 95.4 & 93.3 & 96.3 \\
Tang et al. & 97.7 & 96.2 & 98.3 & 97.3 & 95.0 & 98.4 & 97.6 & 96.1 & 98.2 \\
\hline
Ours - Seq. & \textcolor{blue}{\textbf{98.9}} & \textcolor{blue}{\textbf{98.1}} & \textcolor{blue}{\textbf{99.2}} & \textcolor{red}{\textbf{97.8}} & \textcolor{red}{\textbf{96.8}} & 98.3 & \textcolor{red}{\textbf{98.0}} & \textcolor{red}{\textbf{96.9}} & \textcolor{red}{\textbf{98.5}} \\
Ours - Par. & \textcolor{red}{\textbf{99.0}} & \textcolor{red}{\textbf{98.3}} & \textcolor{red}{\textbf{99.3}} & \textcolor{blue}{\textbf{97.7}} & 96.0 & \textcolor{red}{\textbf{98.6}} & \textcolor{blue}{\textbf{97.9}} & \textcolor{red}{\textbf{96.9}} & \textcolor{blue}{\textbf{98.3}} \\
\bottomrule
\end{tabular} 
\label{tab: MTL-Cls-RETOUCH}
\end{table}

\begin{table}[!h]\small
\caption{Segmentation performance of MTL methods on the RETOUCH dataset. The best and second-best results are respectively highlighted in \textcolor{red}{\textbf{red}}, \textcolor{blue}{\textbf{blue}}.}
\setlength{\tabcolsep}{6pt}
\centering
\begin{tabular}{l|cc|cc|cc}
\toprule
\multirow{2}{*}{Method} & \multicolumn{2}{c|}{Cirrus} & \multicolumn{2}{c|}{Spectralis} & \multicolumn{2}{c}{Topcon} \\
    & JI & DSC & JI & DSC & JI & DSC \\
\midrule
He et al. \cite{he2020multi} & 69.1 & 80.0 & 66.4 & 77.5 & 61.2 & 73.3 \\
Gende et al. \cite{gende2022end} & 69.8 & 80.6 & 67.1 & 78.1 & 60.5 & 73.3 \\
Hervella \cite{hervella2022end} & 68.0 & 78.9 & 67.4 & 78.0 & 60.0 & 72.6 \\
Tang et al. \cite{tang2023transformer} & 69.1 & 79.6 & 67.9 & 78.0 & 62.5 & 74.3 \\
\hline
Ours - Seq. & \textcolor{blue}{\textbf{73.4}} & \textcolor{blue}{\textbf{83.0}} & \textcolor{blue}{\textbf{70.4}} & \textcolor{blue}{\textbf{80.3}} & \textcolor{blue}{\textbf{64.9}} & \textcolor{blue}{\textbf{76.3}} \\
Ours - Par. & \textcolor{red}{\textbf{73.9}} & \textcolor{red}{\textbf{83.4}} & \textcolor{red}{\textbf{71.6}} & \textcolor{red}{\textbf{81.4}} & \textcolor{red}{\textbf{65.7}} & \textcolor{red}{\textbf{76.9}} \\
\bottomrule
\end{tabular} 
\label{tab: MTL-Seg-RETOUCH}
\end{table}

\begin{table}[!t]\small
\caption{Segmentation and classification performances on the ISIC 2017 dataset. The best and second-best results are respectively highlighted in \textcolor{red}{\textbf{red}}, \textcolor{blue}{\textbf{blue}}.}
\setlength{\tabcolsep}{8pt}
\centering
\begin{tabular}{l|cc|ccc}
\toprule
\multirow{2}{*}{Method} & \multicolumn{2}{c|}{Segmentation} & \multicolumn{3}{c}{Classification}\\
& JI & DSC & Acc & Sen & Spe\\
\midrule
He et al. \cite{he2020multi} & 75.0 & 84.1 & 83.2 & 71.4 & 87.8 \\
Gende et al. \cite{gende2022end} & 75.5 & 84.4 & 83.0 & 66.5 & 87.0 \\
Hervella et al. \cite{hervella2022end} & 70.6 & 80.3 & 79.3 & 59.3 & \textcolor{red}{\textbf{89.3}} \\
Tang et al. \cite{tang2023transformer} & 73.7 & 82.4 & 83.1 & 74.0 & 87.7 \\
\hline
Ours - Sequential & \textcolor{red}{\textbf{78.6}} & \textcolor{red}{\textbf{86.6}} & \textcolor{red}{\textbf{84.6}} & \textcolor{red}{\textbf{75.7}} & \textcolor{blue}{\textbf{89.1}} \\
Ours - Parallel & \textcolor{blue}{\textbf{78.4}} & \textcolor{blue}{\textbf{86.5}} & \textcolor{blue}{\textbf{83.6}} & \textcolor{blue}{\textbf{74.3}} & 88.3 \\
\bottomrule
\end{tabular} 
\label{tab: MTL_ISIC}
\end{table}

\begin{figure}[!t]
    \centering
    \includegraphics[width=\columnwidth]{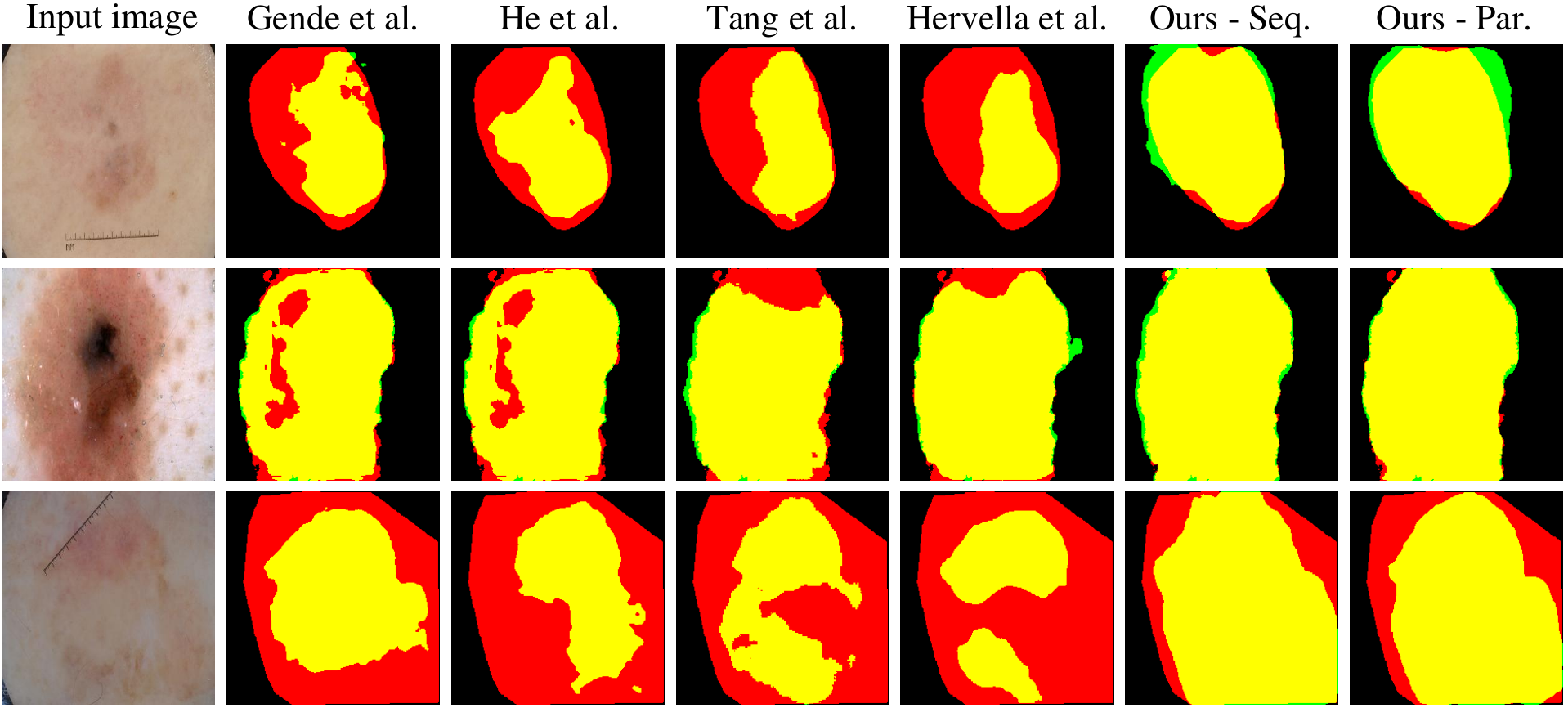}
    \caption{Segmentation visualization of MTL methods on the ISIC 2017 dataset. Yellow, red, and green indicate true positive, false negative, and false positive, respectively.}
    \label{fig: Results_MTL_ISIC}
\end{figure}

\textbf{Results on the ISIC 2017 dataset:} Our comprehensive experiments on the ISIC 2017 dataset encompass both skin lesion classification and segmentation. In the classification task, our method with the sequential ResFormer design achieves 84.6\% accuracy and 75.7\% sensitivity, outperforming other MTL approaches. For segmentation, our parallel variant achieves an impressive 78.6\% JI and 86.6\% DSC, as detailed in Table \ref{tab: MTL_ISIC}. These results surpass the previous SOTA method \cite{gende2022end}, showing improvements of 3.1\% in JI and 2.2\% in DSC. Figure \ref{fig: Results_MTL_ISIC} provides a visual representation of our qualitative results, illustrating the precision of our segmentation approach. This consistent superiority across various datasets underscores the effectiveness and robustness of our method.

\subsection{Comparison with existing STL methods}
We conduct experiments to investigate the effectiveness of our method in comparison to various STL models. Our comparisons include CNN-based and hybrid methods, as well as deep learning models specifically developed for each dataset for both medical image classification and segmentation tasks. To ensure a fair comparison, we strictly follow the original settings described in the papers and employ open-source codes to reproduce the results on two benchmark datasets.

\begin{figure*}[!h]
    \centering
    \includegraphics[width=0.8\textwidth]{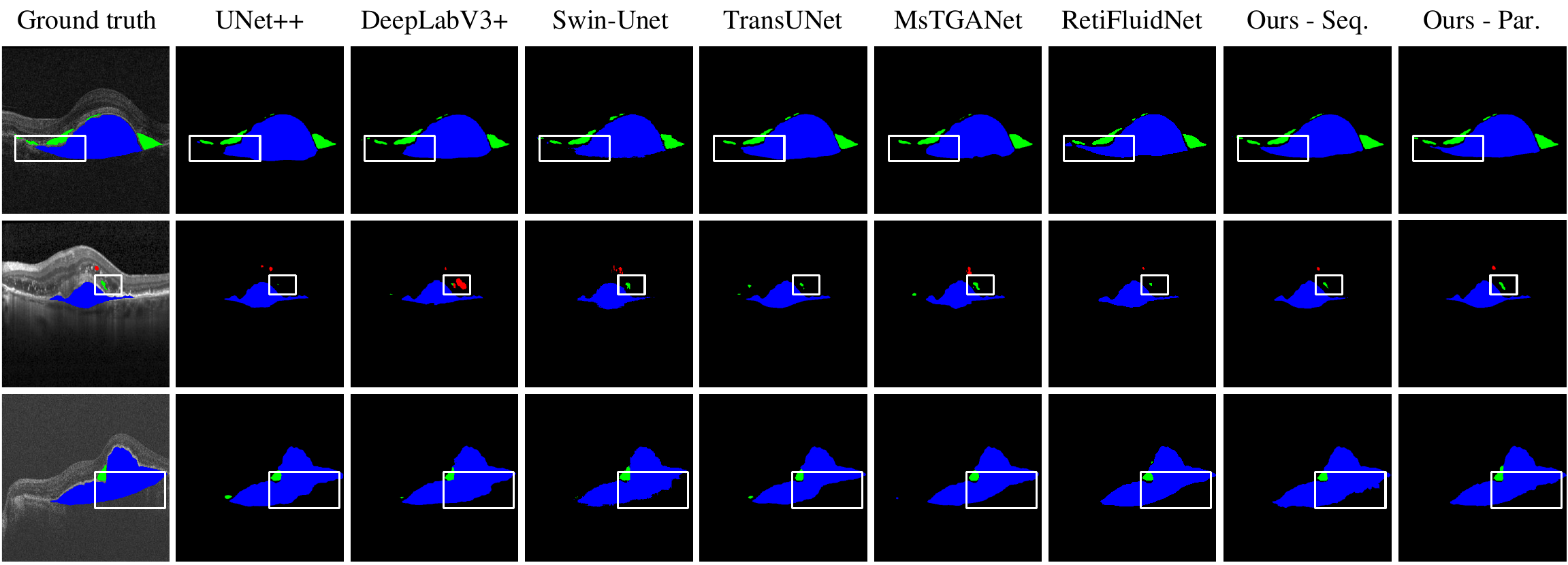}
    \caption{Qualitative segmentation comparison with other STL methods on the RETOUCH dataset. Red, green, and blue indicate intra-retinal fluid, sub-retinal fluid, and pigment epithelial detachment, respectively.}
    \label{fig: STL_RETOUCH}
\end{figure*}

\textbf{Results on the RETOUCH dataset:} The quantitative results for the classification task show that our method surpasses other STL models in terms of accuracy, sensitivity, and specificity, showcasing superior performance across all metrics as shown in Table \ref{tab: STL-Cls-RETOUCH}. In terms of segmentation, the proposed method respectively outperforms best-performing models in three OCT machines by 1.8\%, 2.0\%, and 1.4\% in DSC, as shown in Table \ref{tab: STL-Seg-RETOUCH}. Unlike TransUNet, which incorporates a few transformer layers on features from the last level of the encoder, our approach unifies ResNet34 and Transformer blocks within each encoder block at every level. This integration enables the capture of both local and global features, which are critical for effective segmentation, thereby leading to superior performance. Furthermore, Figure \ref{fig: STL_RETOUCH} shows the results of ours and other STL methods. As highlighted in the white box, our method consistently delivers better results across various fluid scenarios including large and small areas.

\begin{table}[!h]\small
\caption{Classification comparison between ours and STL methods on the RETOUCH dataset. The best and second-best results are respectively highlighted in \textcolor{red}{\textbf{red}}, \textcolor{blue}{\textbf{blue}}.}
\setlength{\tabcolsep}{3pt}
\centering
\begin{tabular}{l|ccc|ccc|ccc}
\toprule
\multirow{2}{*}{Method} & \multicolumn{3}{c|}{Cirrus} & \multicolumn{3}{c|}{Spectralis} & \multicolumn{3}{c}{Topcon} \\
 & Acc & Sen & Spe & Acc & Sen & Spe & Acc & Sen & Spe \\
\midrule
InceptionV3 & 97.6 & 96.3 & 98.1 & 97.4 & \textcolor{blue}{\textbf{96.4}} & 97.9 & 97.7 & 95.7 & \textcolor{blue}{\textbf{98.5}} \\
EfficientNet & 98.5 & 97.6 & 98.8 & 97.1 & 95.6 & 97.9 & 97.7 & 96.0 & 98.4 \\
SwinT & 98.4 & 97.6 & 98.8 & 97.7 & 96.2 & \textcolor{blue}{\textbf{98.5}} & 96.8 & 94.6 & 97.7 \\
OCTNet & 97.9 & 96.6 & 98.4 & 96.8 & 95.8 & 97.3 & 97.9 & 96.1 & \textcolor{red}{\textbf{98.7}} \\
MSL-SRN & 98.0 & 96.5 & 98.6 & 97.2 & 96.1 & 97.8 & \textcolor{red}{\textbf{98.0}} & \textcolor{blue}{\textbf{96.6}} & 95.5 \\
\hline
Ours - Seq. & \textcolor{blue}{\textbf{98.9}} & \textcolor{blue}{\textbf{98.1}} & \textcolor{blue}{\textbf{99.2}} & \textcolor{red}{\textbf{97.8}} & \textcolor{red}{\textbf{96.8}} & 98.3 & \textcolor{red}{\textbf{98.0}} & \textcolor{red}{\textbf{96.9}} & \textcolor{blue}{\textbf{98.5}} \\
Ours - Par. & \textcolor{red}{\textbf{99.0}} & \textcolor{red}{\textbf{98.3}} & \textcolor{red}{\textbf{99.3}} & \textcolor{blue}{\textbf{97.7}} & 96.0 & \textcolor{red}{\textbf{98.6}} & \textcolor{blue}{\textbf{97.9}} & \textcolor{red}{\textbf{96.9}} & 98.3 \\
\bottomrule
\end{tabular} 
\label{tab: STL-Cls-RETOUCH}
\end{table}

\begin{table}[!h]\small
\caption{Segmentation comparison between ours and STL methods on the RETOUCH dataset. The best and second-best results are respectively highlighted in \textcolor{red}{\textbf{red}}, \textcolor{blue}{\textbf{blue}}.}
\setlength{\tabcolsep}{5.5pt}
\centering
\begin{tabular}{l|cc|cc|cc}
\toprule
\multirow{2}{*}{Method} & \multicolumn{2}{c|}{Cirrus} & \multicolumn{2}{c|}{Spectralis} & \multicolumn{2}{c}{Topcon} \\
    & JI & DSC & JI & DSC & JI & DSC \\
\midrule
UNet++ \cite{zhou2018unet++} & 70.8 & 80.6 & 69.0 & 78.7 & 63.6 & 75.2 \\
DeepLabV3+ \cite{chen2018encoder} & 71.5 & 81.2 & 67.7 & 77.7 & 63.0 & 74.7 \\
Swin-Unet \cite{cao2022swin} & 65.5 & 76.2 & 67.1 & 77.3 & 55.8 & 68.5 \\
TransUNet \cite{chen2021transunet} & 71.8 & 81.6 & 67.5 & 77.4 & 63.8 & 75.3 \\
MsTGANet \cite{wang2021mstganet} & 70.8 & 80.8 & 69.3 & 79.4 & 63.9 & 75.5 \\
RetiFluidNet \cite{rasti2022retifluidnet} & 71.9 & 81.6 & 69.5 & 79.3 & 63.1 & 74.6 \\
HiFormer \cite{heidari2023hiformer} & 67.5 & 77.5 & 65.0 & 75.0 & 59.3 & 71.1 \\
HTC-Net \cite{tang2024htc} & 68.5 & 79.0 & 68.5 & 78.7 & 60.4 & 72.8 \\
\hline
Ours - Seq. & \textcolor{blue}{\textbf{73.4}} & \textcolor{blue}{\textbf{83.0}} & \textcolor{blue}{\textbf{70.4}} & \textcolor{blue}{\textbf{80.3}} & \textcolor{blue}{\textbf{64.9}} & \textcolor{blue}{\textbf{76.3}} \\
Ours - Par. & \textcolor{red}{\textbf{73.9}} & \textcolor{red}{\textbf{83.4}} & \textcolor{red}{\textbf{71.6}} & \textcolor{red}{\textbf{81.4}} & \textcolor{red}{\textbf{65.7}} & \textcolor{red}{\textbf{76.9}} \\
\bottomrule
\end{tabular} 
\label{tab: STL-Seg-RETOUCH}
\end{table}

\begin{table}[!t]\small
\caption{Classification and segmentation comparison between ours and STL methods on the ISIC 2017 dataset. The best and second-best results are respectively highlighted in \textcolor{red}{\textbf{red}}, \textcolor{blue}{\textbf{blue}}.}
\setlength{\tabcolsep}{8pt}
\centering
\begin{tabular}{l|ccc|cc}
\toprule
\multirow{2}{*}{Method} & \multicolumn{3}{c|}{Classification} & \multicolumn{2}{c}{Segmentation} \\
             & Acc & Sen & Spe & JI & DSC \\
\midrule
InceptionV3 \cite{szegedy2016rethinking}  & 83.4 & 73.2 & \textcolor{blue}{\textbf{88.5}} & - & - \\
EfficientNet \cite{tan2019efficientnet}  & 82.1 & 69.7 & 88.3 & - & - \\
ViT \cite{dosovitskiy2010image}  & 78.9 & 62.0 & 87.3 & - & - \\
SwinT \cite{liu2021swin}  & 83.0 & 73.3 & 87.8 & - & - \\
\hline
UNet++ \cite{zhou2018unet++} & - & - & - & 76.5 & 84.9 \\
DeepLabV3+ \cite{chen2018encoder} & - & - & - & 76.1 & 84.6 \\
Swin-Unet \cite{cao2022swin} & - & - & - & 70.7 & 80.3 \\
TransUNet \cite{chen2021transunet} & - & - & - & 76.6 & 85.0 \\
autoSMIM \cite{wang2023autosmim} & - & - & - & 77.9 & 85.7 \\
EIU-Net \cite{yu2023eiu} & - & - & - & 77.1 & 85.5 \\
HiFormer \cite{heidari2023hiformer} & - & - & - & 76.6 & 84.8 \\
HTC-Net \cite{tang2024htc} & - & - & - & 77.2 & 85.4 \\
\hline
Ours - Sequential & \textcolor{red}{\textbf{84.6}} & \textcolor{red}{\textbf{75.7}} & \textcolor{red}{\textbf{89.1}} & \textcolor{red}{\textbf{78.6}} & \textcolor{red}{\textbf{86.6}} \\
Ours - Parallel & \textcolor{blue}{\textbf{83.6}} & \textcolor{blue}{\textbf{74.3}} & 88.3 & \textcolor{blue}{\textbf{78.4}} & \textcolor{blue}{\textbf{86.5}} \\
\bottomrule
\end{tabular} 
\label{tab: STL_Cls_ISIC}
\end{table}

\textbf{Results on the ISIC 2017 dataset:} Next, we perform experiments on the ISIC 2017 dataset to compare our method with other STL models for each task. In the classification task, our method outperforms other STL approaches across three evaluation metrics including accuracy, sensitivity, and specificity as shown in Table \ref{tab: STL_Cls_ISIC}. For segmentation, our proposed method demonstrates superior performance compared to SOTA approaches, as detailed in Table \ref{tab: STL_Cls_ISIC}. Specifically, our method incorporating sequential ResFormer blocks achieves an impressive 78.6\% IoU and 86.6\% DSC, surpassing the previous benchmark set by the autoSMIM \cite{wang2023autosmim} method, with improvements of 0.7\% IoU and 0.9\% DSC. Figure \ref{fig: Results_MTL_ISIC} provides a visual representation of our qualitative results, highlighting the precision of our segmentation approach. These consistent improvements across different tasks and datasets highlight the robustness and effectiveness of the proposed method.


\begin{figure*}[!t]
    \centering
    \includegraphics[width=0.8\textwidth]{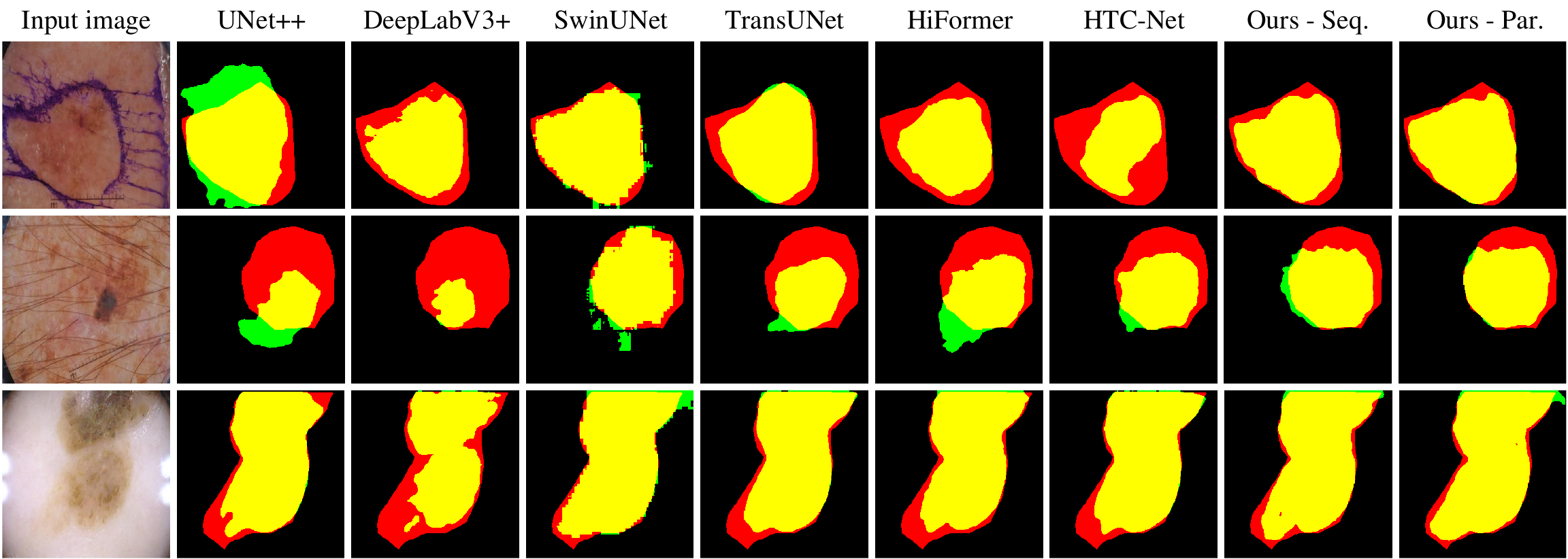}
    \caption{Qualitative segmentation comparison with STL methods on the ISIC 2017 dataset. Yellow, red, and green indicate true positive, false negative, and false positive, respectively.}
    \label{fig7}
\end{figure*}

\begin{table}[!h]\small
\setlength{\tabcolsep}{5pt}
      \centering
        \caption{Effectiveness of multi-task learning. The best results are highlighted in \textbf{bold}.}
        \begin{tabular}{l|c|cc|ccc}
        \toprule
        \multirow{2}{*}{Method} & \multirow{2}{*}{Machine} & \multicolumn{2}{c|}{Segmentation} & \multicolumn{3}{c}{Classification} \\
             &  & JI & DSC & Acc & Sen & Spe \\
        \midrule
        \multicolumn{7}{c}{Sequential} \\
        \hline
        \multirow{3}{*}{Cls. only} & Cirrus & - & - & 98.1 & 97.1 & 98.6 \\
                  & Spectralis & - & - & 97.3 & 95.8 & 98.1 \\
                & Topcon & - & - & 97.8 & 96.3 & 98.5 \\
        \hline
        \multirow{3}{*}{Seg. only} & Cirrus & 73.0 & 82.5 & - & - & - \\
                  & Spectralis & 69.6 & 79.4 & - & - & - \\
                & Topcon & 64.3 & 75.5 & - & - & - \\
            \hline
        \multirow{3}{*}{Both} & Cirrus &\textbf{73.4} & \textbf{83.0} & \textbf{98.9} & \textbf{98.1} & \textbf{99.2} \\
            & Spectralis & \textbf{70.4} & \textbf{80.3} & \textbf{97.8} & \textbf{96.8} & \textbf{98.3} \\
            & Topcon & \textbf{64.9} & \textbf{76.3} & \textbf{98.0} & \textbf{96.9} & \textbf{98.5} \\
        \midrule
        \multicolumn{7}{c}{Parallel} \\
        \hline
        \multirow{3}{*}{Cls. only} & Cirrus & - & - & 98.6 & 98.1 & 98.7 \\
                  & Spectralis & - & - & 97.3 & 95.8 & 98.0 \\
                & Topcon & - & - & 97.8 & 95.6 & \textbf{98.4} \\
        \hline
        \multirow{3}{*}{Seg. only} & Cirrus & 73.3 & 83.0 & - & - & - \\
                  & Spectralis & 70.6 & 80.0 & - & - & - \\
                & Topcon & 64.8 & 76.2 & - & - & - \\
            \hline
        \multirow{3}{*}{Both} & Cirrus &\textbf{73.9} & \textbf{83.4} & \textbf{99.0} & \textbf{98.3} & \textbf{99.3} \\
            & Spectralis & \textbf{71.6} & \textbf{81.4} & \textbf{97.7} & \textbf{96.0} & \textbf{98.6} \\
            & Topcon & \textbf{65.7} & \textbf{76.9} & \textbf{97.9} & \textbf{96.9} & 98.3 \\
        \bottomrule
        \end{tabular}
\label{tab: ablation_mtl}
\end{table}

\subsection{Ablation studies}
To thoroughly assess our method, we perform extensive experiments to evaluate the effectiveness of multi-task learning and the impact of the DFE module on both the sequential and parallel variants. As shown in Table \ref{tab: ablation_mtl}, MTL considerably enhances performance in both segmentation (Seg.) and classification (Cls.) tasks across both variants of our method. These results underscore the synergistic benefits of jointly learning related tasks, i.e. segmentation and classification.

Regarding the DFE module, Table \ref{tab: ablation_dfe} showcases the positive impact of the DFE module on improving the segmentation performance in two variants of our method. In particular, the sequential variant witnesses performance improvements of 0.5\%, 0.5\%, and 0.3\% in JI across three data sources while the parallel variant achieves gains of 0.7\%, 1.3\%, and 0.6\% in JI with the integration of the DFE module in its structure. This comprehensive evaluation underlines the DFE module's effectiveness in enhancing segmentation performance.

\begin{table}[!h]\small
\setlength{\tabcolsep}{8pt}
      \centering
        \caption{Effectiveness of DFE module in segmentation task. The best results are highlighted in \textbf{bold}.}
        \begin{tabular}{l|c|cc|cc}
        \toprule
        \multirow{2}{*}{Method} & \multirow{2}{*}{Machine} & \multicolumn{2}{c}{Sequential} & \multicolumn{2}{c}{Parallel} \\
         & & JI & DSC & JI & DSC \\
        \midrule
        \multirow{3}{*}{w/o DFE} & Cirrus & 72.9 & 82.7 & 73.2 & 82.9 \\
                                & Spectralis & 69.9 & 79.6 & 70.9 & 80.4 \\
                                & Topcon & 64.6 & 76.1 & 65.1 & 76.4 \\
        \hline
        \multirow{3}{*}{w/ DFE} & Cirrus & \textbf{73.4} & \textbf{83.0} & \textbf{73.9} & \textbf{83.4} \\
                                & Spectralis & \textbf{70.4} & \textbf{80.3} & \textbf{71.6} & \textbf{81.4} \\
                                & Topcon & \textbf{64.9} & \textbf{76.3} & \textbf{65.7} & \textbf{76.9} \\
        \bottomrule
        \end{tabular}
\label{tab: ablation_dfe}
\end{table}

\section{Conclusion}
In this paper, we introduce a novel MTL framework designed to effectively integrate both segmentation and classification tasks for medical image analysis. Our method uses novel ResFormer blocks that combine the advantages of CNNs and Transformers, enabling the model to capture both local and global features crucial for precise medical image interpretation. Furthermore, we propose a dilated feature enhancement (DFE) module, which includes three parallel convolution branches with different dilation rates. This module aggregates multi-scale contextual information for the segmentation decoder, thereby enhancing the detection of lesions of various sizes. Our comprehensive experiments on the RETOUCH and ISIC 2017 datasets demonstrated that the proposed method consistently outperforms SOTA single-task and multi-task learning approaches. Specifically, our model demonstrates significant improvements in both JI and DSC for segmentation tasks, as well as better accuracy, sensitivity, and specificity in classification tasks. 

Although our method has achieved notable performance, it still relies on fully annotated datasets with both classification and segmentation labels. In future work, we plan to employ semi-supervised or weakly-supervised learning to reduce the need for extensive labeling and enhance performance in scenarios where segmentation labels are sparse or incomplete.


\bibliographystyle{IEEEtran}
\bibliography{refs}

\end{document}